\documentclass[11pt,manuscript]{article}
\usepackage{epsfig}
\usepackage{graphicx}
\usepackage{epsfig}

\begin{document}
\title{Monte Carlo Simulation of Surface De-alloying of Au/Ni(110)}
\author{W.\ Fan$^{1}$ \footnote{corresponding to W.Fan,
        email:~~fan@theory.issp.ac.cn} and X.\ G.\ Gong$^{2,1}$ \\
 $^{1}$ Laboratory for Computational Materials Science, \\
 Institute of Solid State Physics, Hefei Institutes of Physical Science, \\
 Chinese Academy of Sciences, 230031-Hefei, P. R. China \\
  $^{2}$Surface Physics Laboratory (National Key Laboratory) and Department of Physics, \\
  Fudan University, 200433-Shanghai, P. R. China}

\date{\today}
\maketitle

\bigskip
\bigskip
\bigskip
\bigskip

\begin{abstract}
 \narrower{\small
 Based on BFS model and using Monte Carlo simulation we confirm
 the de-alloying in immiscible Au/Ni(110) system. The critical
 Au coverage 0.4(ML) when de-alloying happens is consistent with
 experiments. At the same time our simulation show that the
 structural phase transition will lead to the saturation of the
 number of alloying Au atoms.
}
\end{abstract}

\begin{center} {\bf \Large I.  Introduction } \end{center}

 Surface alloying is an important phenomenon in hetero-growth of
 surfaces [1-4]. From Scanning Tunnelling Microscopy
 (STM) observations, gold atoms mix with copper atoms and form the Au-Cu
$c(2\times2)$ structure in the  Au/Cu(110)
 system~\cite{Chambliss1}; gold atoms mix with the silver atoms at next atomic layer below the surface
 layer for Au/Ag(110)~\cite{Rousset1}. Theoretically, the equilibrium structures of
 surface alloys can be obtained from total-energy calculations~\cite{Daw1,Chan1}.
 Surface alloying of immiscible systems, such as Au/Ni(110)~\cite{Nielsen1},
 (Na,K)/Al(111)[10-13], Ag/Pt(111)~\cite{Roder1},
 Sb/Ag(111)~\cite{Oppo1}, is an astonishing phenomenon because it is difficult to mix
 different kind atoms in bulk for these elements. Thus, we can synthesis some new and
 novel compounds on surface which are difficult to synthesis in bulk material.
 A few theoretical models have been presented to explain
 surface alloys in immiscible systems such as Effective-Medium
 Theory (EMT)~\cite{Nielsen1}, Tersoff Theory~\cite{Tersoff1} and BFS Model~\cite{Bozzolo1,Bozzolo2}.
 Especially, the BFS model~\cite{Bozzolo1,Bozzolo2,Bozzolo3} has been successfully used to study
 the phenomenon of surface alloying in immiscible systems [29-36].

 Recent experiments~\cite{Nielsen2} on the Au/Ni(110) system have
 shown that when Au coverage is above 0.4ML a
 new phenomenon, "de-alloying", takes place. Au atoms, having already
 alloyed, segregate again and form complex one-dimensional chains
 perpendicular to the closed-packed direction. A density
 functional calculation has shown the interesting properties of these
 one-dimensional chains~\cite{Pampuch1}.
 In this work we study the alloying and de-alloying of Au/Ni(110).
 Generally Au atoms are immiscible in Ni crystal. The experimental
 finding of Au atoms alloying to the Ni(110) surface inspired the
 interests of immiscible systems. Based on the BFS model, several
 theoretical calculations and computer simulations have
 explained the experimental phenomenon of alloying. In this
 work, using Monte Carlo methods, we find that the BFS model can
 successfully explain both alloying and de-alloying of Au/Ni(110).
 The results show that when Au coverage increases above
 0.40ML, de-alloying occurs, which is consistent with experimental observations.

 \begin{center} {\bf \Large II.  BFS Model and Monte Carlo Methods } \end{center}

 The Construction of the BFS model is based on the concept of Equivalent Crystal
 Theory (ECT)~\cite{Smith1,Smith2}, that is, the formation of defects in a perfect
 crystal equivalent to the changes of lattice constant of a perfect crystal. If the
 lattice constant of the equilibrium perfect crystal is $a_{0}$, ECT
 theory requires that there exists an equivalent lattice with lattice
 constant $a \neq a_{0}$ whose energy is equal to the energy of
 the deformed lattice ( the crystal after forming defects).
 The energy of the equivalent lattice can be calculated from the
 universal energy relation [23-26]
 in terms of the change of that lattice constant which is obtained from the ECT equation.

 The BFS model generalizes the ECT theory to alloy systems, that is,
 including the chemical effects. The change of the chemical environment
 is also equivalent to a change of lattice constant and the
 corresponding energy can be computed following the same steps as
 the calculation of the energy related to defects in the crystal. Below, we
 briefly describe the BFS model. More details can be found in Ref.~\cite{Bozzolo3}.

 The formation energy of an arbitrary alloy structure is the
 superposition of individual contributions, $\varepsilon_{i}$, of
 the nonequivalent atoms in the alloy,

 \begin{eqnarray}
  \Delta H &=& \sum_{i} \varepsilon_{i}^{S}+g_{i}\varepsilon_{i}^{chem}
 \end{eqnarray}

 \noindent where $\varepsilon_{i}^{S}$ is the strain
 energy, $\varepsilon_{i}^{chem}$ is the chemical energy, and $g_{i}$ is
 a coupling function. The strain energy is calculated from

 \begin{equation}
 \varepsilon_{i}^{S}=E_{C}^{i}F^{*}(a_{i}^{S*}),
 \end{equation}

 \noindent where
 $F^{*}(a^{*})=1- ( 1 + a^{*})e^{-a^{*}}$ is an universal
 scaling function.
 $a_{i}^{S*}=q \frac{a_{i}^{S}-a_{e}^{i}}{l_{i}}$ is the scaled
 change of lattice constant, $a_{i}^{S}$ is the lattice
 constant of the equivalent crystal, and $a_{e}^{i}$ is the lattice constant for
 the perfect lattice. The parameters $\it q $ and $\it l$ describe the structure
 and properties of system~\cite{Smith1}.
 The coupling function is $g_{i}=e^{-a_{i}^{S*}}$. The lattice constant of the equivalent
 crystal, $a_{i}^{S}$, can be obtained from the ECT equation

 \begin{equation}
  NR_{1}^{p}e^{-\alpha R_{1}}+MR_{2}^{p}e^{-[\alpha+(1/\lambda)]R_{2}}
  = \sum_{j}r_{j}^{p}e^{-[\alpha+S(r_{j})]r_{j}}
 \end{equation}

 \noindent where $R_{1}$($R_{2}$) are the distances of first nearest(next nearest)
 neighbor of atom $i$, $N$($M$) is the number of nearest(next nearest)
 neighbors. $p, \alpha, \lambda$ are the screening parameters defined in
 Ref~\cite{Smith1,Bozzolo5}. $S(r_{1})=0$ for nearest
 neighbors and $S(r_{2})=1 /\lambda$ for next nearest neighbors.

 Each atom in the alloy has its own equivalent crystal, whose lattice
 constant is dependent on the environment of the atom. In
 the calculation of the strain energy, $\varepsilon_{i}^{S}$, the structural effects
 are considered and chemical effects are ignored , that is, all
 atoms near atom $i$ are considered to be of the same kind as atom $i$.
 Chemical effects are included in the chemical energy
 of, $\varepsilon_{i}^{chem}$, in which structural effects are ignored.
 The calculation of chemical energy is
 similar to that of the strain energy.  The equivalent lattice for atom $i$
 is obtained from

 \begin{eqnarray}\scriptstyle
  & & NR_{1}^{p_{i}}e^{-\alpha_{i} R_{1}}+MR_{2}^{p_{i}}e^{-[\alpha_{i}+(1/\lambda_{i})]R_{2}} \nonumber \\
  &=&\sum_{k}r_{1}^{p_{i}}e^{-\alpha_{ik}r_{1}}+\sum_{k}r_{2}^{p_{i}}e^{-[\alpha_{ik}+(1/\lambda_{i})]r_{1}}
 \end{eqnarray}

 Compared with Eq.3, the only difference is the appearance of the parameter
 $\alpha_{ik}=\alpha_{ik}+\Delta_{ki}$ for atom $i$ which depends
 on the type of neighboring atoms $k$. If atom $k$ is identical to atom
 $i$, $\Delta_{ik}=\Delta_{ii}=0$, otherwise,$\Delta_{ik} \neq 0$
 which is just the chemical effect. In the BFS model, the chemical
 energy includes only the chemical environment and ignores the
 structural facts, thus, $r_{1}, r_{2}$ are the nearest and next
 nearest neighbor distances in the equilibrium crystal of species $i$ and retain the
 chemical environment of atom $i$. If we have solved the ECT
 equation, the chemical energy can be calculated according to
 $\varepsilon_{i}^{chem}=\varepsilon_{i}^{C}-\varepsilon_{i}^{C_{0}}$.
 The $\varepsilon_{i}^{C_{0}}$ is the energy without chemical
 impurities($\Delta_{ik}=0$). The calculation of $\varepsilon_{i}^{C}$
 and $\varepsilon_{i}^{C_{0}}$ are summarized below

 \begin{equation}\scriptstyle
  \varepsilon_{i}^{C}=\gamma E_{C}^{i}F^{*}(a_{i}^{C*})~~~~~~~~~~~~~~~~~¡¡
  \varepsilon_{0_{i}}^{C}=\gamma_{_{0}} E_{C}^{i}F^{*}(a_{i}^{C_{0}*})
 \end{equation}

 \begin{equation}\scriptstyle
  a_{i}^{C^{*}}    =\frac{q(a_{i}^{C}    -a_{i}^{e})}{l_{i}}~~~~~~~~~~~~~~~~~¡¡
  a_{i}^{C^{*}_{0}}=\frac{q(a_{i}^{C_{0}}-a_{i}^{e})}{l_{i}}
 \end{equation}

 \begin{equation}\scriptstyle
  \gamma = \left\{
  \begin{tabular}{cc}
   1 & $a_{i}^{C^{*}}\geq 0 $ \\
  -1 & $a_{i}^{C^{*}} < 0 $ \\
  \end{tabular}
  \right. ~~~~~and~~~~~ \gamma_{_{0}} = \left\{
 \begin{tabular}{cc}
  1 & $a_{i}^{C^{*}_{0}}\geq 0 $ \\
 -1 & $a_{i}^{C^{*}_{0}} < 0 $ \\
 \end{tabular}
 \right.
 \end{equation}

 We have used Monte Carlo methods to simulate the surface alloying
 and de-alloying based on the BFS model.
 In the simulation, we calculate the energy for every
 configuration when sampling the phase space of the system under study.
 Rejection or acceptance of a configuration is based on the Metropolis
 criterion~\cite{Metropolis1}.
 The key of a Monte Carlo simulation is the calculation of the total energy.
 For the BFS model, $\Delta H = \sum_{i} \varepsilon_{i}$.
 $\varepsilon_{i}$ is dependent on the environment of atom $i$. If we
 consider a perfect lattice that includes only one kind of atoms, the
 environment of atom $i$ is the nearby configurations of atom $i$
 which can be described as $(n_{i},m_{i})$ where $n_{i}$($ m_{i}$) is the number of nearest
 (next nearest) neighbors of atom $i$.
 So, $\Delta H = \sum_{i} \varepsilon(n_{i},m_{i})$. Because we only
 consider the nearest and next nearest neighbors of an atom, the possible configurations
 are very limited for a typical crystal (fcc,bcc,hcp and diamond structure,
 for fcc metal the number is 91). All
 configurations for every atom are cast into this group of the
 91 configurations. If having calculated energies of the 91
 configurations before entering the main loop of Monte Carlo,
 in the main loop we only need to find which one of the 91 configurations corresponds
 to the current configuration. The total energy can be calculated
 from the equations given above. So we calculate energies only 91 times in our
 simulation which is independent on the number of atoms under study.
 Exactly defining nearby configurations for every atom
 is the main task in our simulation.
 For alloy systems, the available nearby configurations are large,
 but every configuration is only calculated one time.
 In our simulation, atoms move in a 3D
 lattice. The atom $i$ jump from site $m$ to $n$, if the site $n$
 is empty, the atom $i$ occupies the site $n$. If the site $n$ has
 been occupied by another atom $j$ with the different kind of atom $i$,
 the atom $i$ exchanges position with atom $j$ or the jump is
 rejected.

 \begin{figure}
 \begin{center}  \includegraphics[width=5.0in]{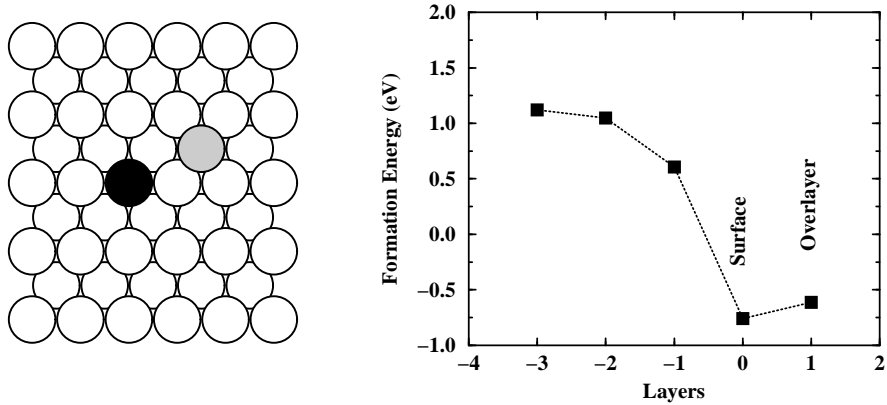}
 \caption{Formation energies of an Au atom absorbed on surface and
 substituted other Ni atom in or below surface layer;
 (0) is indicative of the surface layer; (1) the over-layer;
 (-1),(-2),(-3) and (-4) the one, two, three and four below the surface
 respectively. The black sphere is the alloying Au atom, gray spheres denote
 Ni atom substituted by Au and white spheres represent the Ni atoms of the substrate}
 \label{fig1}
 \end{center}
 \end{figure}

\begin{center} {\bf \Large III. Results and Discussion} \end{center}

 The simulation cell includes 5 atomic layers, and each
 layer contains 280 atoms. The number of substrate atoms of the
 cell is 1400. We generally deposited Au atoms with
 coverage from 0 to 1ML, or from 0 to 280 Au atoms in order to study
 the alloying and de-alloying of the Au/Ni(110) system. Two atomic
 layers at the bottom are fixed and periodical boundary conditions are applied to the
 directions parallel to the surface. In our simulation, the direct
 exchanging of different kinds of atoms is allowed if an atom moves
 to a site already occupied by a different kind of atom. The
 sweep number of the Monte Carlo simulation is 50000 or 100000.

 Fig.~\ref{fig1} shows the formation energy when a Au atom substitutes for a Ni
 atom in different layers. The left panel shows the Au atom
 substituting for a Ni atom in the surface layer. From this figure we find that
 the lowest formation energy corresponds to an Au atom substituting for a Ni
 atom in the surface layer. Our results also indicate that Au atoms do not alloy into the
 Ni crystal, proof of immiscibility of the Au/Ni system.
 A similar result, that is, surface alloying of immiscible systems, has been
 reported by other authors~\cite{Bozzolo2}.

 \begin{figure}
 \begin{center}  \includegraphics[width=2.5in]{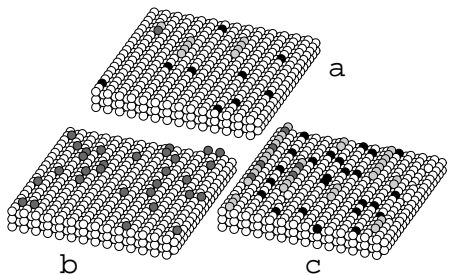}
 \caption{Surface structures at low coverage (T=300K). (a) 0.036ML; (b)
 Initial configuration at 0.1(ML);(c) Final configuration at 0.1(ML).
 Ni atoms in and below the surface layer are denoted with white spheres;
 Ni atoms on the surface represent with light gray spheres;
 Au atoms alloying into the surface layer are denoted with black spheres;
 Au atoms on the surface represent with dark gray spheres.}
 \label{fig2}
 \end{center}
 \end{figure}

 Fig.~\ref{fig2} shows the surface structure at low Au coverage at room
 temperature T=300K. Fig.~\ref{fig2}(a) shows the relaxed surface structure at
 0.036 ML Au coverage. We find from this figure that all Au atoms
 are alloying into the surface layer except an Au atom that is still on
 the surface. The Ni atoms on the surface, which are substituted by
 alloying Au atoms, form two Ni chains of 4 Ni atoms. The
 remaining Au atom on the surface forms a dimer with a Ni atom. By
 increasing Au coverage to 0.1ML, the length of Ni chains also increases
 with increasing number of alloying Au atoms. Alloying
 atomic chains also appear, and their lengths also increase.
 Some Au atoms, which have no chance allying into surface layer,
 form the alloying chains with Ni atoms. The chance of forming alloying chains
 is larger than that at lower coverage (Fig.~\ref{fig2}(a)). Our results are
 in close agreement with experimental observations. An important
 difference compared with experiment is the smaller numbers of Ni
 dimers alloying into the surface layer, contrary to experiment
 where most Ni atoms are paired.

 \begin{figure}
 \begin{center}  \includegraphics[width=2.5in]{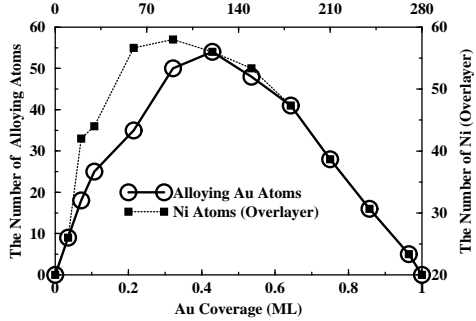}
 \caption{The changes in the number of Au atoms alloying into the surface layer
 and in the number of Ni atoms on the surface formed
 by substituting Ni atoms with Au atoms in the substrate.}
 \label{fig3}
 \end{center}
 \end{figure}

 In the experiment at higher Au coverage, the de-alloying
 phenomenon is found with increasing Au coverage up to
 0.40ML~\cite{Nielsen2}. The number of alloying Au atoms does not
 increase, on the contrary, it decreases with increasing Au coverage.
 On the surface, one-dimensional Au chains with complicated structure
 perpendicular to the closed packed direction are found. We now study the
 change of surface structure thoroughly from low coverage to high
 coverage. The Monte Carlo simulation sweeps 100000 steps for each
 coverage at T=300K. The results are shown in Fig.~\ref{fig3}.
 From this figure we find that the number of Au atoms alloying into
 the surface layer decreases if the Au coverage increases continuously
 above 0.4ML. Our results are consistent with the experimental
 observation. The critical coverage for de-alloying
 is exactly 0.40, the experimental value. From Fig.~\ref{fig3}
 we also find that the number of Ni atoms on the surface is larger than the number of
 alloying Au atoms when Au coverage is lower than the critical
 value, which indicates that there are vacancies on the surface at
 low Au coverage. If $n$ Ni atoms jump from the prefect surface layer
 onto the surface, $n$ vacancies (unoccupied sites ) will be left in
 the surface layer; If $m$ Au atoms alloy into the surface layer and
 occupy $m$ unoccupied sites, the number of vacancies in the surface
 layer will decrease to $n-m$. If $n=m$, all sites in the
 surface layer are occupied and no vacancies are in the surface layer.
 When the de-alloying happens, some Au atoms return to the surface
 and at the same time some Ni atoms return to the surface layer. From
 the figure, we also find the number of Ni atoms on the surface is
 equal to the number of Au atoms alloying into surface layer. This
 means that a Au atom only substitutes for a Ni atom and all the sites
 of surface layer are occupied without vacancies.

 \begin{figure}
 \begin{center}  \includegraphics[width=2.5in]{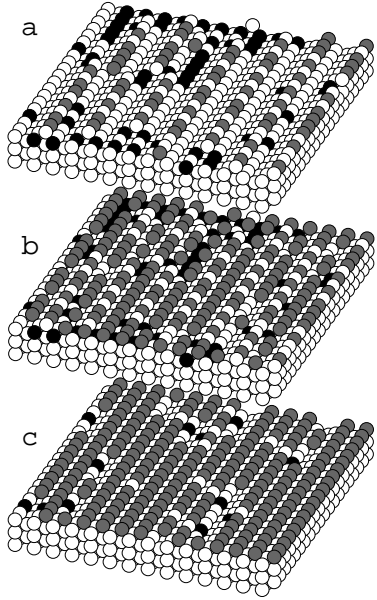}
 \caption{The de-alloying of Au/Ni(110)(T=300K).
 (a) Surface structure at 0.40ML Au coverage; (b) 0.8ML Au coverage
 forms by depositing another 0.40ML Au atoms on Figure(a);
 (c) Surface structure relaxing from (b) by Monte Carlo Methods.
 Ni atoms at, below and on the surface layer are denoted with white
 spheres; Au atoms alloying into the surface layer represent with
 black spheres; Au atoms on surface (Over-layer) are denoted with
 dark gray spheres.}
 \label{fig4}
 \end{center}
 \end{figure}

\begin{figure}
 \begin{center}  \includegraphics[width=2.5in]{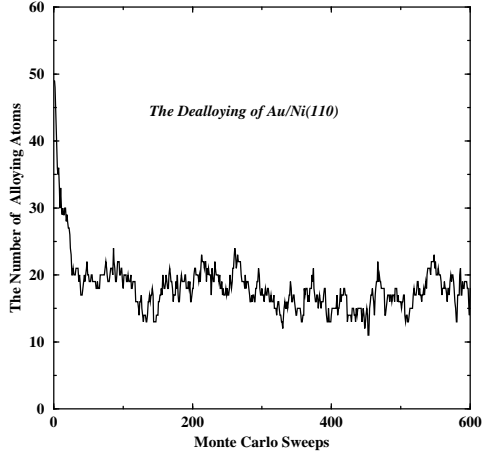}
 \caption{The change of the number of Au atoms as a function of
  Monte Carlo steps (from Fig.4(b) to Fig.4(c))}
 \label{fig5}
 \end{center}
 \end{figure}

 Fig.~\ref{fig4}(a) shows the equilibrium structure at 0.4ML Au coverage
 after a Monte Carlo relaxation of 100000 steps, and Fig.~\ref{fig4}(b) shows the
 surface structure after randomly depositing another 0.40ML Au on
 the already relaxed surface (Fig.~\ref{fig4}(a)). Now we
 use the configuration Fig.~\ref{fig4}(b)
 (with coverage 0.8ML) as the new initial structure and relax
 another 100000 Monte Carlo steps. The structure in Fig.~\ref{fig4}(b) is
 unstable and relaxes to another equilibrium structure. We find, in
 the final surface structure, most of Au atoms return to surface.
 Only a small part of Au atoms is inserted in the alloying surface
 layer. Fig.~\ref{fig5} shows the change in the number of Au atoms alloying
 into the surface layer with Monte Carlo steps. We find the number
 decreases rapidly and approaches the equilibrium value within
 only 100 Monte Carlo steps.

\begin{figure}
 \begin{center}\includegraphics[width=2.5in]{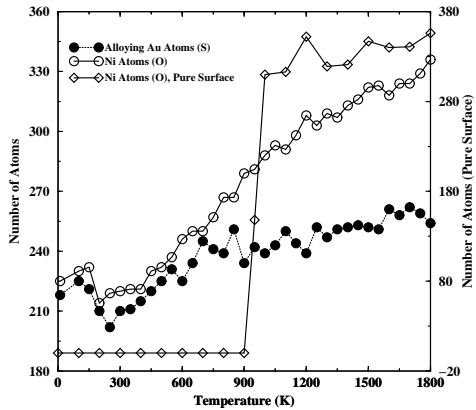}
 \caption{ The changes in the number of Au atoms alloying into the surface and
 the number of Ni atoms staying on the surface with increasing temperature.
 Ni atoms on the surface, Au atoms alloying into the surface and Ni atoms on a pure Ni(110)
 surface are denoted with solid disks, open circles and diamonds respectively.}
 \label{fig6}
 \end{center}
 \end{figure}

 The motion of atoms is generally faster at higher temperatures, an
 atom has more chance to exchange with other atoms. Thus we expect
 that the degree of alloying will increase with temperature. We
 study the temperature effects of surface alloying using a larger cell
 which includes 5 atomic layers and contains 1120 Ni atoms in each
 atomic layer. We also deposit randomly 450 Au atoms on the
 surface. The total number of atoms in the cell is 6050.
 Fig.~\ref{fig6}
 shows the change of the number of Au atoms alloying into the surface
 layer and the change of the number of Ni atoms on surface. We find
 the number of alloying Au atoms increase with increasing
 temperature until it saturates above T=800K.

 From Fig.~\ref{fig6} we find that the number of Ni atoms on the surface
 increases linearly with increasing number of
 alloying Au atoms as long as the temperature is below 800K. When
 the temperature increases above 800K, the number of Ni atoms on the surface also increases
 linearly at the same time that the number of alloying Au atoms
 approaches saturation. The figure shows that the number of Ni
 atoms on the surface is generally larger than the number of alloying
 Au atoms. The difference becomes larger as the temperature increases
 above 800K. This is indicative of a change of surface structure.
 The number of vacancies increases when the temperature rises above 800K. The
 decrease of proportion of substrate Ni atoms of the surface layer
 leads to the decrease of the chance of exchanging with Au atoms on
 surface, which prevents the increase of the number of alloying Au
 atoms. The structural change of the surface is the direct reason of
 the saturation of the number of alloying Au atoms when the temperature
 increases above 800K. The argument is confirmed by a following
 simulation of a pure Ni(110) surface. With increasing temperature,
 some Ni atoms will jump onto the surface because of the thermal motion
 of atoms. When the temperature is above a critical temperature the
 surface will undergo a structural change or structural phase
 transition. The change in the number of Ni atoms on the surface is also shown
 in Fig.~\ref{fig6}. We find that when the temperature increases
 to about 900K, the
 number of Ni atoms on the surface increases abruptly from about 190 to
 350. The jump in the number of Ni atoms on the surface indicates the
 structural change of the surface. The transition temperature, 900K, is
 close to the temperature when the number of alloying Au atoms
 approaches saturation.

\begin{center} {\bf \Large IV.  Conclusion } \end{center}

 In conclusion, Monte Carlo simulations confirm the de-alloying
 in the immiscible Au/Ni(110) system.  The critical Au coverage when
 de-alloying happens is also consistent with experiment. Furthermore,
 our simulation shows that the structural phase
 transition will lead to the saturation of the number of alloying
 Au atoms.

\begin{center}{\bf ACKNOWLEDGMENTS}\end{center}

 W.Fan thanks D.Y.Sun for useful discussion, This work was
 supported by Chinese Academy of Sciences under KJCX2-SW-W11;
 Center for computational science, Hefei Institutes of Physical
 Science, Chinese Academy of Sciences;
 Nature Science Foundation of China, the Special funds for major
 state basic research and GAS projects,

\newpage

\end{document}